# Effect of Twist Angle on Structural, Electronic and Magnetic Properties of Carbon Nano Hybrids: A DFT Study

Amrish Sharma[1], Sandeep Kaur[1], Hitesh Sharma[2], Neha Kapila[3], V. K. Jindal[3], Vladimir Bubanja[4,5] and Isha Mudahar[6*]

[1]Department of Physics, Punjabi University, Patiala.
[2]Department of Physics, IKG Punjab Technical University, Kapurthala.
[3]Department of Physics, Panjab University, Chandigarh.
[4]Measurement Standards Laboratory of New Zealand, Callaghan Innovation, PO Box 31310, Lower Hutt, 5040, Wellington, New Zealand.
[5]The Dodd-Walls Centre for Photonic and Quantum Technologies, University of Otago, 730 Cumberland Street, Dunedin, 9016, New Zealand.
[6]Department of Basic and Applied Sciences, Punjabi University, Patiala.
*dr.ishamudahar@gmail.com

**Abstract: -** Density functional calculations of hybrids consisting of a single wall carbon nanotube and a graphene nanoribbon have been performed. We consider the dependence of the structural, electronic and magnetic properties of the hybrids on the twist angle between their subunits. We calculated the binding energies, pyramidalization angles, Mulliken charge, and HOMO-LUMO gaps as functions of the twist angle. We find that, owing to the asymmetrical spin density distributions of their subunits, the hybrids have finite magnetic moments.

**Introduction**

Varying the twist angle between low dimensional nanomaterials has recently emerged as a fundamentally new direction in device engineering. This new approach is in a stark contrast to the conventional methods where the variability of electronic properties of materials is attained by changing their chemical composition. Current intense research in this field is in large part due to the surprising discovery of superconductivity in twisted bilayer graphene (TBG) at magic angles (for which Fermi velocity vanishes) [1]. One of the guiding ideas of twistronics has been to create different crystal cell sizes and symmetries by introducing Moiré patterns [2]. The consequences of such symmetries have been investigated within the continuum model for TBG, parametrized by interlayer coupling for AA and AB stacking [3]. It was shown that in the limit of vanishing coupling for AA stacking, the Hamiltonian of the system acquires chiral symmetry (a unitary particle-hole symmetry) which leads to flattening of the entire lowest band in the electronic band structure. The eigenfunctions of this Hamiltonian can be mapped onto the lowest Landau level wavefunctions on a torus [4], providing a topological explanation for the origin of absolutely flat band solution of this model of TBG [3]. In TBG devices, rotated at 1.1°±0.1°, Mott insulator behaviour has been observed at half filling of the superlattice's first miniband [5], whereas unconventional superconductivity was observed at slightly higher and lower electrostatic doping [1]. The above extremes in conductivity have also been observed at 15% smaller twist angle [6]. Recent studies have further revealed states with non-zero Chern numbers and orbital magnetism [7]. In addition to the above findings for small twist angles, interesting phenomena were discovered at larger angles, such as quasi-crystalline structure at 30° [8]. Properties of such incommensurate heterostructures with rotational but no translational symmetry cannot be analysed by using Bloch's theorem.

Bilayer structures formed by other materials offer a wealth of opportunities for further explorations. For example, large Moiré patterns are formed at small twist angles between graphene and hexagonal boron nitride owing to their closely matched lattice constants. Experimental technique has been developed [9], which enables on-demand control of the orientation between the layers forming such heterostructures. This has enabled a study of the frictional forces between the layers as well as four-point resistance measurement of graphene while changing the relative orientation of the hexagonal boron nitride layer.

Recent advances in atomically precise fabrication of ultra-narrow graphene nanoribbons (GNRs) [10], as well as chirality-controlled synthesis of single-walled carbon nanotubes (SWCNTs) [11], offer great promise for developing all carbon (C) electronic devices integrating these components. In light of developments in twistronics, it is of interest to consider the properties of hybrids formed by GNRs and SWCNTs. In this paper, we study the effect of the twist angle between a GNR and a SWCNT on structural, electronic and magnetic properties of their hybrids by employing ab-initio calculations based on the density functional theory.

**Computational Details**

We employ the Spanish Initiative for Electronic Simulation with Thousands of Atoms (SIESTA) computational code [12-15]. The calculations are carried out by using the generalized gradient approximation (GGA) that implements Perdew, Burke and Ernzerhof (PBE) exchange correlation functional [16]. Core electrons are replaced by non-relativistic, norm-conserving pseudopotentials generated by improved Troullier-Martins scheme [17]. The valence electrons are described by using the linear combination of numerical pseudo-atomic orbitals of Sankey-Niklewski type [18] and generalized for multiple-ζ and polarization functions. Split valence double-ζ polarized (DZP) basis set with an energy cutoff of 250–300Ry has been used. The structures are obtained by minimizing of the total energy by using Hellmann-Feynman forces, including Pulay-like corrections, until the residual forces acting on each atom were smaller than 0.03eV/Å. The calculated total energies and equilibrium structures were tested for convergence with respect to the energy cutoff.

Test calculations have been performed on armchair graphene nanoribbon (AGNR) and zigzag graphene nanoribbon (ZGNR) with configuration consisting of 60 carbon atoms passivated with hydrogen atoms along the length axis. The calculated C-C bond lengths on armchair and zigzag edges were 1.38Å and 1.42Å, respectively; away from the edges the bond length was 1.44Å. The calculated HOMO-LUMO gaps of AGNR and ZGNR were 1.52eV and 0.67eV, respectively, which was in a good agreement with the available experimental and theoretical results [19-24].

To investigate the interaction strength, the binding energy ($E_b$) of a hybrid is computed as,

$$E_b = E_{HYBRID} - E_{SWCNT} - E_{GNR}$$

where $E_{HYBRID}$, $E_{SWCNT}$ and $E_{GNR}$ are the total energies of the SWCNT-GNR hybrid, single wall carbon nanotube and graphene nanoribbon respectively.

**Results and Discussion**

**Structural properties**

We consider two types of hybrid systems. One of them, with components shown in Figure 1, consists of a (5,5) armchair carbon nanotube (ACNT) and a $N_y$=4 zigzag graphene nanoribbon (ZGNR); the corresponding lengths and widths of these two components are chosen to be almost similar. Another considered system consists of an (8,0) zigzag carbon nanotube (ZCNT), of length 19.88Å and diameter 6.26Å, and a $N_y$=6 armchair graphene nanoribbon (AGNR), of length 18.47Å and width 6.19Å. Hydrogen passivation of carbon dangling bonds for both above systems is done along the open ends of CNTs and along the longer edges of GNRs. In both cases GNR is placed 1.50Å above the SWCNT and rotated around its centre by a twist angle (θ) of 0°, 30°, 60° and 90° respectively.

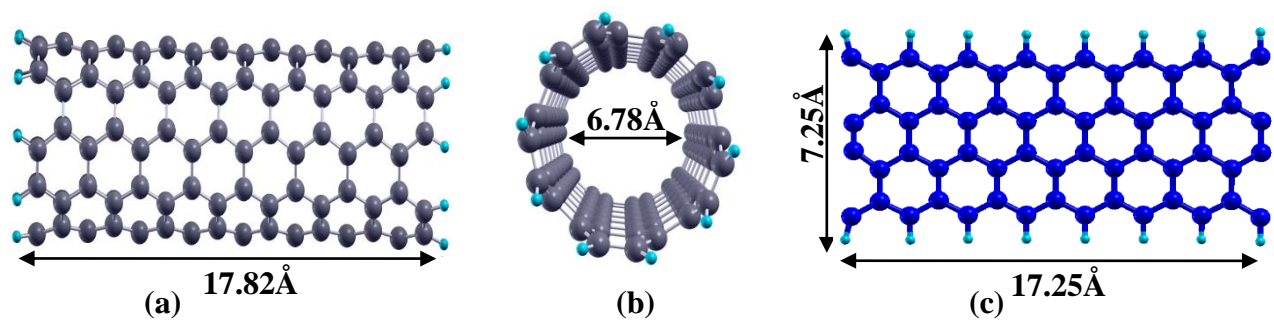

**Figure 1. Ball and stick model for SWCNT and GNR; (a, b) ACNT (c) ZGNR.**

The considered configurations of an ACNT-ZGNR hybrid are shown in Figure 2. In order to study the effects of the stacking arrangements, for θ=0° we investigate three cases regarding the position of graphene nanoribbon in the horizontal plane above the carbon nanotube. In the first case, shown in Figure 2 (a), the central rows of hexagons of graphene nanoribbon and carbon nanotube are aligned on top of each other. In analogy to bulk graphite, we denote this arrangement as AA stacking. In the second case, graphene nanoribbon is translated along its longer axis by $a\sqrt{3}/2$ (where $a$ is the C-C bond length). This arrangement, shown in Figure 2 (b), is denoted as AB[1] stacking. In the third case, graphene nanoribbon is translated to the left by the distance $a$, with respect to the first case. This arrangement, which is shown in Figure 2 (c), is denoted as AB[2] stacking. Similarly, at θ=60°, we consider AA and AB² stacking cases, as shown in Figure 2 (e) and Figure 2 (f), respectively. The binding energies and the interlayer separation between the subunits for all cases are listed in Table 1. We obtain that the interlayer separation in AA stacking is larger than in AB stacking, hence the former is less stable, similar to the case of bilayer graphene [25-27]. The most stable structure is for θ=0° and AB¹ stacking arrangement, for which we obtain the binding energy -3.26eV.

**Table 1. The Binding Energy ($E_b$) and Interlayer Separation (D) of all SWCNT-GNR configurations.**

| ACNT-ZGNR | | | | ZCNT-AGNR | | | |
|---|---|---|---|---|---|---|---|
| θ (°) | Stacking | $E_b$ (eV) | D (Å) | θ (°) | Stacking | $E_b$ (eV) | D (Å) |
| 0 | AA | -3.01 | 3.15 | 0 | AA | -2.95 | 3.12 |
| | AB¹ | -3.26 | 3.02 | | AB² | -3.28 | 3.04 |
| | AB² | -3.04 | 3.06 | 30 | | -2.73 | 3.29 |
| 30 | | -2.66 | 3.09 | 60 | AA | -1.71 | 3.07 |
| 60 | AA | -1.91 | 3.06 | | AB¹ | -1.83 | 2.99 |
| | AB² | -2.08 | 3.00 | | AB² | -1.79 | 3.01 |
| 90 | | -1.66 | 3.11 | 90 | | -1.50 | 3.00 |

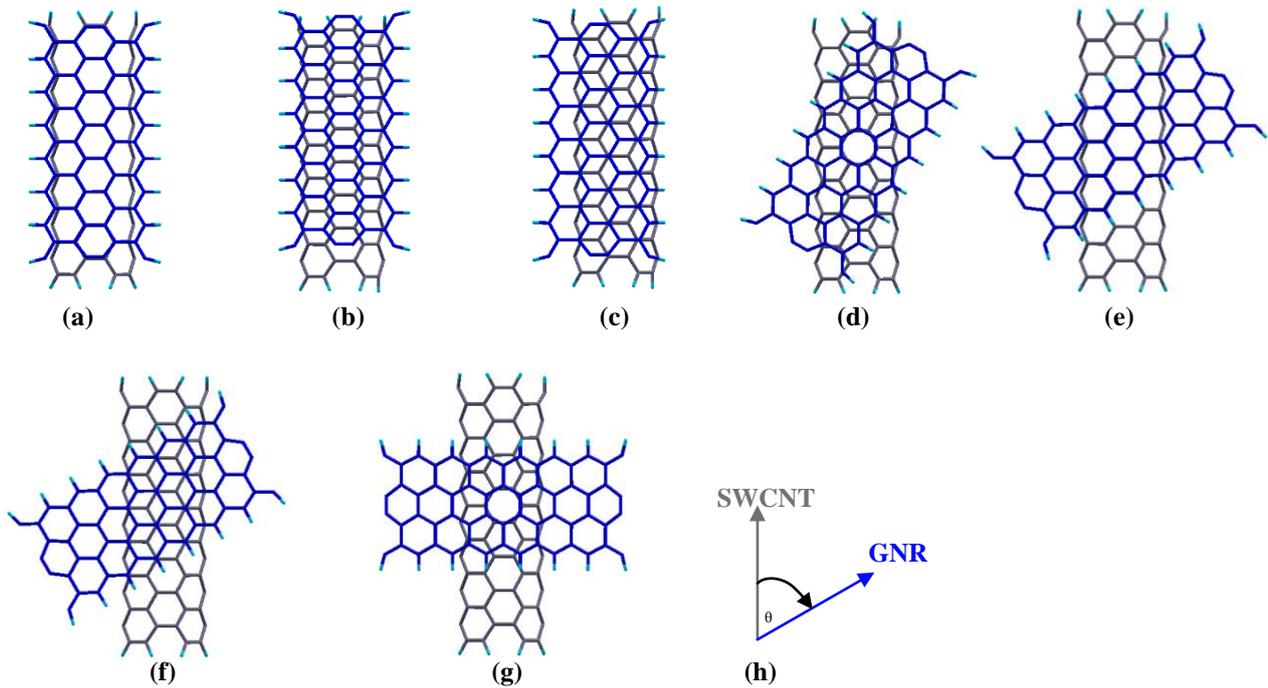

(a)   (b)   (c)   (d)   (e)

(f)   (g)   (h)

**Figure 2.** Stick model for ACNT-ZGNR configurations; (a) 0°/AA (b) 0°/AB¹ (c) 0°/AB² (d) 30° (e) 60°/AA (f) 60°/AB² (g) 90°. Plot (h) indicates the angle θ between the long axes of ACNT and ZGNR.

To investigate the interaction between the ZCNT and the AGNR the configurations with the same twist angles (0⁰, 30⁰, 60⁰ and 90⁰) are considered. These configurations, with the corresponding stacking orders are displayed in Figure 3. After structural optimization, we found the most stable structure at θ=0° for $AB^2$ stacking arrangement having binding energy -3.28eV. Ball and stick models for relaxed structures of ACNT-ZGNR (0°/$AB^1$) and ZCNT-AGNR (0°/$AB^2$) are shown in Figures 4(a) and 4(b), respectively. Moreover, in GNR and SWCNT hexagonal rings, no significant change is seen in the bond length and bond angle of the internal carbon atoms, but on the peripheral carbons the bond angle changes up to 2°.

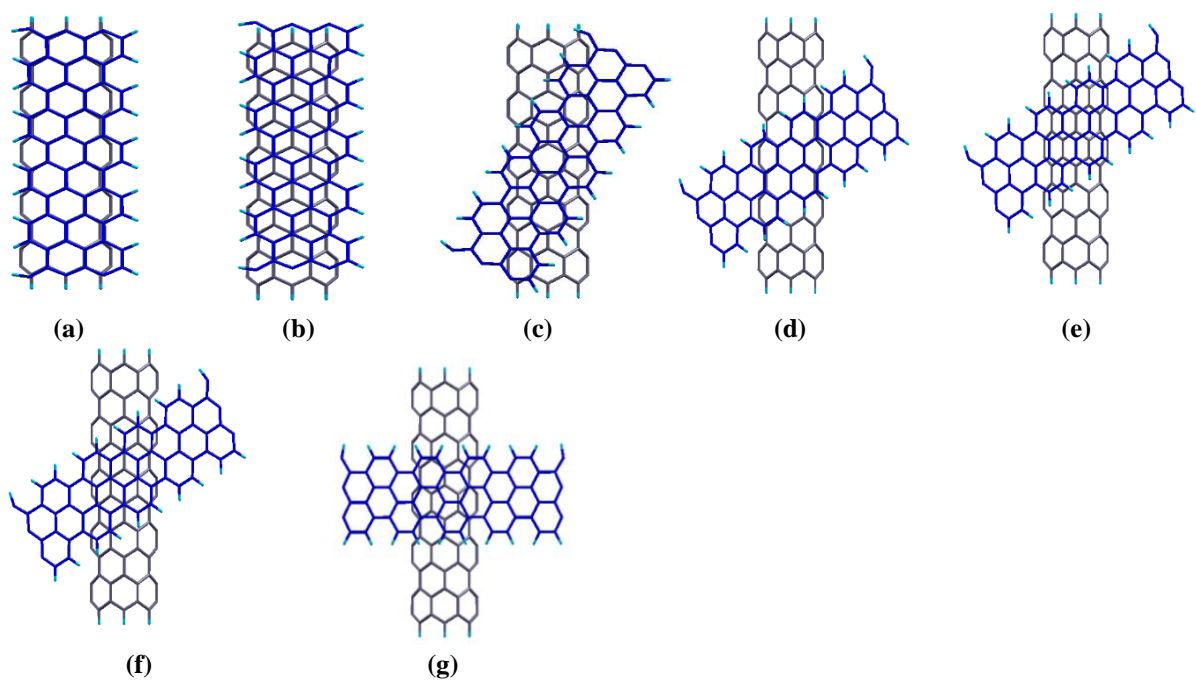

**Figure 3. Stick model for ZCNT-AGNR configurations; (a) 0°/AA (b) 0°/$AB^2$ (c) 30° (d) 60°/AA (e) 60°/$AB^1$ (f) 60°/$AB^2$ (g) 90°.**

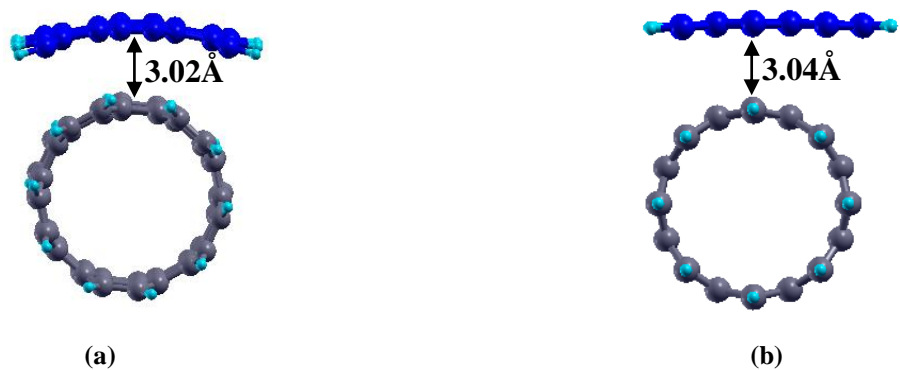

**Figure 4. Ball and stick model for relaxed structure; (a) ACNT-ZGNR (0°/$AB^1$) (b) ZCNT-AGNR (0°/$AB^2$).**

The reactivity of the hybrid is investigated by calculating the pyramidalization angle ($\beta_P$) by using the π-orbital axis vector method [28,29]. The pyramidalization angle, defined by $\beta_P = \beta_{\sigma\pi} - \pi/2$, describes the deviation due to curvature of an ideally planar arrangement of a graphene, where a carbon atom and its three neighbors are located in the same plane. The π-orbital axis vector is defined as the vector that makes the same angles $β_{σπ}$ to the three σ bonds at the central carbon atom to its three neighbors. The calculated pyramidalization angle of nanotube and ribbon in the hybrids after full structural optimization is shown in Table 2. The values of pyramidalization angles for isolated armchair and zigzag SWCNT are 6.12° and 6.40° respectively, while for armchair and zigzag edge GNR the angles are almost negligible as they are flat. However, the comparison of pyramidalization angles of isolated GNR, SWCNT and their hybrids shows that there is a significant change in the pyramidalization angle after formation of the hybrids.

Table 2. Pyramidalization Angle ($β_P$) of SWCNT and GNR in relaxed structures of the hybrid.

| \multicolumn{3}{c}{ACNT-ZGNR} | | | \multicolumn{3}{c}{ZCNT-AGNR} | | |
|---|---|---|---|---|---|
| θ (°) | Stacking | $β_P$ (°) ACNT, ZGNR | θ (°) | Stacking | $β_P$ (°) ZCNT, AGNR |
| 0 | AA | 5.60, 1.66 | 0 | AA | 5.89, 1.05 |
|  | AB$^1$ | 5.69, 1.56 |  | AB$^2$ | 5.98, 0.68 |
|  | AB$^2$ | 5.74, 0.40 | 30 |  | 5.84, 1.38 |
| 30 |  | 5.60, 0.52 | 60 | AA | 5.86, 1.03 |
| 60 | AA | 5.54, 0.05 |  | AB$^1$ | 5.84, 0.84 |
|  | AB$^2$ | 5.69, 0.79 |  | AB$^2$ | 5.76, 0.69 |
| 90 |  | 5.57, 0.16 | 90 |  | 5.58, 0.54 |

**Electronic Properties**

The electronic properties of the hybrids are investigated in terms of the Mulliken charge transfer (|e|), energy difference between the spin-up (↑) and spin-down (↓) highest occupied molecular orbital (HOMO) and lowest unoccupied molecular orbital (LUMO) level and electron density of states (eDOS) plot.

The interaction of GNR with SWCNT changes the position of the Fermi level of nanotubes, resulting in the charge transfer between the subunits. When the twist angle θ=0°, charge on a nanotube (ACNT or ZCNT) in AA stacking increases by 0.02|e| and 0.11|e|, respectively, whereas the value lies between 0.03|e| and 0.08|e| in AB (AB$^1$ or AB$^2$) stacking. The configurations obtained in 30° show very small increase in charge on both the nanotubes. The behaviour get reversed in 60$^0$ and $90^0$ orientation respectively, where all the configurations show charge transfer below -0.06|e|, except $60^o$/AA stacking in ZCNT-AGNR; this configuration shows the maximum of transferred charge (-1.13|e|) amongst all considered configurations. The charge transfer in most stable structure ($0^o$/AB$^1$) of ACNT-ZGNR is 0.04|e|, whereas ($0^o$/AB$^2$) for ZCNT-AGNR is 0.08|e|.

The difference in energy of HOMO and LUMO levels for spin-up and spin-down electronic states for all configurations is shown in Table 3. For ACNT-ZGNR hybrid there is a variation of HOMO-LUMO gap for spin-up states of about a factor of two between the lowest ($0^o$/AA) and the highest ($90^o$) values, with no significant change for spin-down states. The situation is reversed for ZCNT-AGNR.

In order to further elucidate the origin and change in the electronic properties of hybrids, the spin-up and spin-down eDOS of the most stable configurations are plotted in Figure 5. The plots show a noticeable change in the eDOS of hybrids with respect to isolated subunits. The unequal distributions of spin-up and spin-down states indicate finite magnetic moments of the hybrids. For ACNT-ZGNR hybrid both spin-up and spin-down DOS are vanishing at the Fermi level, while for ZCNT-AGNR there is a finite DOS only for the spin-up states.

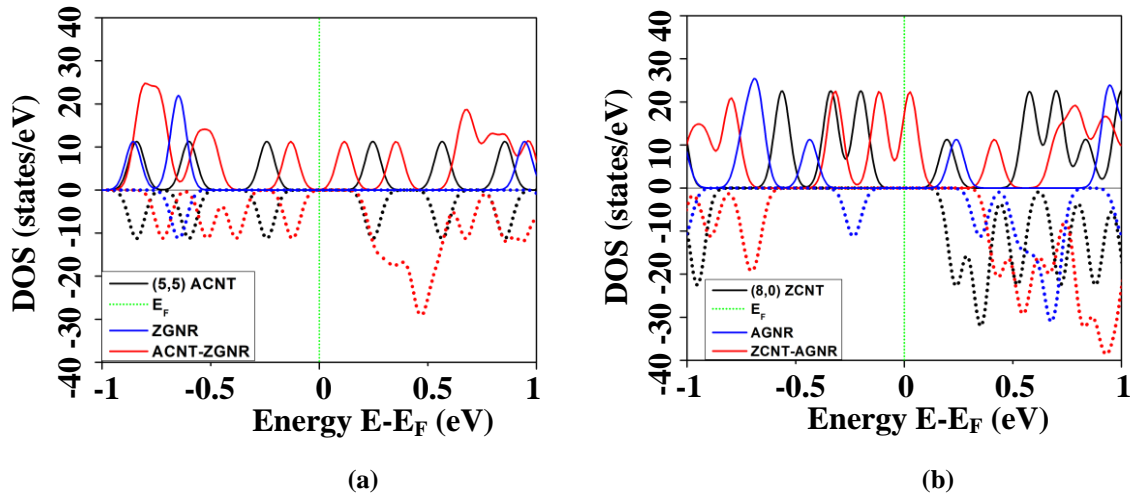

Figure 5. The spin electron density of states (eDOS) for most stable hybrid; (a) ACNT-ZGNR ($0^o$/AB$^1$) (b) ZCNT-AGNR ($0^o$/AB$^2$). Solid lines are for spin up and dashed lines for spin down states.

**Magnetic Properties**

The magnetic properties of the hybrids are investigated in terms of the total magnetic moment (TMM), local magnetic moment (LMM) and the spin density plots. The TMMs, calculated from the difference in total spin-up and spin-down electron densities, for all the considered configurations are summarized in Table 3.

Table 3. The Total Magnetic Moments (TMM) and the HOMO-LUMO (H-L) gaps (eV) for spin up (↑) and spin down (↓) states in the considered hybrid structures.

| | ACNT-ZGNR | | | | ZCNT-AGNR | | | |
|---|---|---|---|---|---|---|---|---|
| θ (º) | Stacking | TMM ($\mu_B$) | H-L (↑) | H-L (↓) | θ (º) | Stacking | TMM ($\mu_B$) | H-L (↑) | H-L (↓) |
| 0 | AA | 8.79 | 0.23 | 0.39 | 0 | AA | 14.66 | 0.32 | 1.08 |
| | $AB^1$ | 8.02 | 0.25 | 0.38 | | $AB^2$ | 15.54 | 0.37 | 1.05 |
| | $AB^2$ | 8.02 | 0.24 | 0.36 | 30 | | 14.60 | 0.38 | 1.12 |
| 30 | | 7.93 | 0.30 | 0.44 | 60 | AA | 12.13 | 0.39 | 0.57 |
| 60 | AA | 7.74 | 0.40 | 0.49 | | $AB^1$ | 14.08 | 0.39 | 1.08 |
| | $AB^2$ | 7.80 | 0.34 | 0.42 | | $AB^2$ | 13.24 | 0.38 | 1.10 |
| 90 | | 9.56 | 0.47 | 0.43 | 90 | | 13.90 | 0.40 | 1.09 |

For an isolated armchair and zigzag nanotubes, we obtain TMMs of about 3.05$\mu_B$ and 6.00$\mu_B$, respectively, while for armchair and zigzag edge nanoribbon we obtain TMM of 8.00$\mu_B$. For a ACNT-ZGNR hybrid the magnitude of TMM is in the range 7.74$\mu_B$ and 9.56$\mu_B$, whereas for ZCNT-AGNR the value lies between 12.13$\mu_B$ to 15.54$\mu_B$. Their most stable configurations show TMM of 8.02$\mu_B$ and 15.54$\mu_B$, respectively. In most stable configuration of ACNT-ZGNR hybrid (0º/$AB^1$), the LMM shows that TMM of tube is 0.07$\mu_B$ which is reduced by antiferromagnetic interactions between carbon atoms of ribbon. The major contribution to TMM is observed at hydrogenated edge carbon atoms of the ribbon, which contributes 87% of TMM. In order to visualize the magnetic ordering, the electron spin distribution of the hybrid is plotted in Figure 6(a); the plot shows an asymmetrical spin charge density on both subunits which result in the magnetic moment of the hybrid.

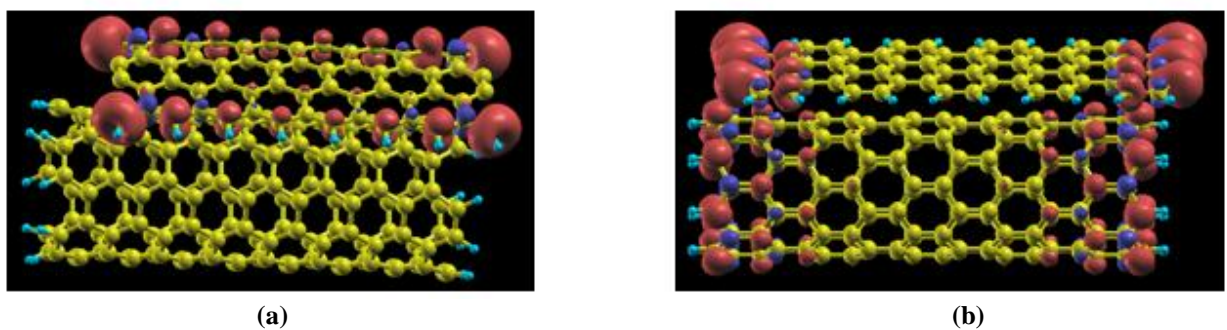

**Figure 6.** Spin density maps for most stable hybrid; (a) ACNT-ZGNR ($0^o$/$AB^1$) (b) ZCNT-AGNR ($0^o$/$AB^2$).

In case of the ZCNT-AGNR hybrid, the TMM of all configurations decreases with increase in the twist angle. The variation of the TMM can be understood in terms of the LMM on the inner and edge carbon atoms. For the most stable configuration of hybrid ($0^o$/$AB^2$), the LMM of nanotube increases by 19% due to interaction with the nanoribbon, whereas the LMM of the ribbon increases by 5% due to interaction. The hydrogenated edge carbon atoms of ZCNT and AGNR contribute to the TMM with 41% and 37%, respectively. The carbon atoms on the zigzag edges along the nanoribbon width contribute 53% to the TMM. Therefore, our calculations suggest that the magnetism in hybrids result from the states along the zigzag edges, which is in agreement with the previous studies [30]. The spin density plot in Figure 6(b) shows that the zigzag edge carbon atoms of the hybrid interact ferromagnetically resulting in the high total magnetic moment.

**Summary & Conclusions**

The effects of the twist angle on structural, electronic and magnetic properties of carbon nanotube-graphene nanoribbon hybrids were considered by using the first-principle calculations. The structural optimization shows that formation of these hybrids is energetically favourable, with most stable structures formed at $0^o$ for AB stacking owing to strong interlayer coupling between the layers. The flat ZGNR changes to curved shape after its interaction with ACNT but no curvature is seen in AGNR-ZCNT hybrid. There are no changes in bond lengths and bond angles of inner carbon atoms in the hexagonal rings of GNR and SWCNT, but the bond angles on the edges change up to $2^o$. The increase in angle of twist decreases the pyramidalization angle of SWCNT and GNR in hybrids, thus making them relatively less reactive.

The interaction between GNR and SWCNT results in the significant change in their electronic and magnetic properties. When twist angle is $0^o$ and $30^o$, the charge on nanotube (ACNT or ZCNT) increases, whereas charge on GNR (AGNR or ZGNR) decreases. The behaviour is reversed in $60^0$ and $90^0$ orientations. All the configurations possess finite HOMO-LUMO gaps. Change in the twist angle between the subunits could be used to tune the HOMO-LUMO gaps of the hybrids. It would be of interest to develop analytic approach, based on theory developed in [31], that would couple the Tomonaga-Luttinger liquid (corresponding to CNT) and tight binding (GNR) Hamiltonians with coupling parametrized by the twist angle.

**Acknowledgements**

The authors are grateful to the New Zealand eScience Infrastructure(NeSI) for providing high performance computing facilities. Amrish Sharma is thankful to DST-SERB, Govt. of India for providing the equipment to carry out the work. Sandeep Kaur is thankful to UGC (University Grant Commission) New Delhi, India for giving financial support. The authors acknowledge the SIESTA group for providing their computational code.

**Data Availability**

The data that support the findings of this study are available from the corresponding author upon reasonable request.

# References

[1] Y. Cao, V. Fatemi, S. Fang, K. Watanabe, T. Taniguchi, E. Kaxiras and P. J. Herrero, Nature **556**, 43 (2018). DOI: **https://doi.org/10.1038/nature26160**.

[2] R. Bistritzer and A.H. MacDonald, Proceedings of the National Academy of Sciences **108(30)**, 12233 (2011). DOI: **https://doi.org/10.1073/pnas.1108174108**.

[3] G. Tarnopolsky, A.J. Kruchkov and A. Vishwanath, Phys. Rev. Lett. **122**, 106405 (2019). DOI: https://doi.org/10.1103/PhysRevLett.122.106405.

[4] F.D.M. Haldane and E.H. Rezayi, Phys. Rev. B **31**, 2529 (1985). DOI: https://doi.org/10.1103/PhysRevB.31.2529.

[5] Y. Cao, V. Fatemi, A. Demir, S. Fang, S.L. Tomarken, J.Y. Luo, J.D.S. Yamagishi, K. Watanabe, T. Taniguchi, E. Kaxiras, R.C. Ashoori and P.J. Herrero, Nature **556**, 80 (2018). DOI: https://doi.org/10.1038/nature26154.

[6] E. Codecido, Q. Wang, R. Koester, S. Che, H. Tian, R. Lv, S. Tran, K. Watanabe, T. Taniguchi, F. Zhang, M. Bockrath and C.N. Lau, Sci. Adv. **5,** eaaw9770 (2019). DOI: 10.1126/sciadv. aaw9770.

[7] X. Lu, P. Stepanov, W. Yang, M. Xie, M.A. Aamir, I. Das, C. Urgell, K. Watanabe, T. Taniguchi, G. Zhang, A. Bachtold, A.H. MacDonald and D.K. Efetov, Nature **574**, 653 (2019). DOI: https://doi.org/10.1038/s41586-019-1695-0.

[8] S.J. Ahn, P. Moon, T.H. Kim, H.W. Kim, H.C. Shin, E.H. Kim, H.W. Cha, S.J. Kahng, P. Kim, M. Koshino, Y.W. Son, C.W. Yang and J.R. Ahn, Science **361**, 782 (2018). DOI: 10.1126/science.aar8412.

[9] R.R. Palau, C. Zhang, K. Watanabe, T. Taniguchi, J. Hone and C.R. Dean, Science **361,** 690 (2018). DOI: 10.1126/science. aat6981.

[10] A. Kimouche, M.M. Ervasti, R. Drost, S. Halonen, A. Harju, P.M. Joensuu, J. Sainio and P. Liljeroth, Nat. Commun. **6**, 10177 (2015). DOI: https://doi.org/10.1038/ncomms10177.

[11] J. Tomada, T. Dienel, F. Hampel, R. Fasel and K. Amsharov, Nat. Commun. **10**, 3278 (2019). DOI: https://doi.org/10.1038/s41467-019-11192-y.

[12] P. Ordejón, E. Artacho, and J.M. Soler, Phys. Rev. B **53**, R10441 (2002). DOI: https://doi.org/10.1103/PhysRevB.53.R10441.

[13] P.D. Sánchez, P. Ordejón, E. Artacho and J.M. Soler, Int. J. Quantum Chem. **65**, 453 (1997). DOI: https://doi.org/10.1002/(SICI)1097-461X(1997)65:5%3C453::AID-QUA9%3E3.0.CO;2-V.

[14] J.M. Soler, E. Artacho, J.D. Gale, A. García, J. Junquera, P. Ordejón and D.S. Portal, J. Phys. Condens. Matter **14**, 2745 (2002). DOI: https://doi.org/10.1088/0953-8984/14/11/302.

[15] Parr R.G. (1980) Density Functional Theory of Atoms and Molecules. In: Fukui K., Pullman B. (eds) Horizons of Quantum Chemistry. Académie Internationale Des Sciences

Moléculaires Quantiques / International Academy of Quantum Molecular Science, vol 3. Springer, Dordrecht. DOI: https://doi.org/10.1007/978-94-009-9027-2_2.

[16] J.P. Perdew, K. Burke, and M. Ernzerhof, Phys. Rev. Lett. **77(18)**, 3865 (1997). DOI: https://doi.org/10.1103/PhysRevLett.77.3865.

[17] L. Kleinman and D.M. Bylander, Phys. Rev. Lett. **48**, 1425 (1982). DOI: https://doi.org/10.1103/PhysRevLett.48.1425.

[18] O.F. Sankey and D.J. Niklewski, Phys. Rev. B **40**, 3979 (1989). DOI: https://doi.org/10.1103/PhysRevB.40.3979.

[19] M. Hammouri and I. Vasiliev, Physica E **89**, 170 (2017). DOI: http://dx.doi.org/10.1016/j.physe.2017.02.019.

[20] I.K. Petrushenko, Journal of Nano- and Electronic Physics **9(3)**, 03018 (2017). DOI: http://doi.org/10.21272/jnep.9(3).03018.

[21] N.M. Díez, A.G. Lekue and E.C. Sanromà, ACS Nano **11(11)**, 11661 (2017). DOI: https://doi.org/10.1021/acsnano.7b06765.

[22] S. Kaur, A. Sharma, H. Sharma, S. Dhiman and I. Mudahar, Int. J. Quantum Chem. **119**, e26019 (2019). DOI: https://doi.org/10.1002/qua.26019.

[23] S. Kaur, H. Sharma, V.K. Jindal, V. Bubanja and I. Mudahar, Physica E **111,** 1 (2019). DOI: https://doi.org/10.1016/j.physe.2019.02.018.

[24] A. Sharma, S. Kaur, H. Sharma and I. Mudahar, Mater. Res. Express **5(6)**, 065032 (2018). DOI: https://doi.org/10.1088/2053-1591/aacb18.

[25] E. Mostaani, N. D. Drummond, and V. I. Fal'ko, Phys. Rev. Lett. **115**, 115501 (2015). DOI: https://doi.org/10.1103/PhysRevLett.115.115501.

[26] I.V. Lebedeva, A.A. Knizhnik, A.M. Popov, Y.E. Lozovik, B.V. Potapkin, Phys Chem Chem Phys. **13(13)**, 5687 (2011). DOI: https://doi.org/10.1039/C0CP02614J.

[27] H. Santos, A. Ayuela, L. Chico, and E. Artacho, Phys. Rev. B **85**, 245430 (2012). DOI: https://doi.org/10.1103/PhysRevB.85.245430.

[28] H.C. Bai, Z. Ying, Y.N. Ni, J.Y. Qiang, Q.W. Ye and H.Y. He, Chinese J. Struct. Chem. **32(5)**, 695 (2013).

[29] S. Niyogi, M.A. Hamon, H. Hu, B. Zhao, P. Bhowmik, R. Sen, M.E. Itkis and R.C. Haddon, Acc. Chem. Res. **5**, 1105 (2002). DOI: https://doi.org/10.1021/ar010155r.

[30] H. Şahin, C. Ataca and S. Ciraci, Phys. Rev. B **81**, 205417 (2010). DOI: https://doi.org/10.1103/PhysRevB.81.205417.

[31] V. Bubanja, and S. Iwabuchi, Phys. Rev. B **79**, 035312 (2009). DOI: https://doi.org/10.1103/PhysRevB.79.035312.